\documentclass[10pt,conference]{IEEEtran}
\IEEEoverridecommandlockouts
\usepackage{cite}
\usepackage{amsmath,amssymb,amsfonts}
\usepackage{algorithmic}
\usepackage{graphicx}
\usepackage{textcomp}
\usepackage{xcolor}
\usepackage{cases}
\usepackage[version=4]{mhchem} 
\usepackage{makecell} 

\begin{document}
\bstctlcite{IEEEexample:BSTcontrol} 

\title{Stochastic Modeling of Biofilm Formation with Bacterial Quorum Sensing \\ 
\thanks{The works of FG and AWE were funded by a Discovery grant from the Natural Sciences and Engineering Research Council of Canada.
}}


\author{\IEEEauthorblockN{Fatih Gulec, Andrew W. Eckford, 
	}
\IEEEauthorblockA{Department of Electrical Engineering and Computer Science, York University, Toronto, Ontario, Canada \\
Email: fgulec@yorku.ca, aeckford@yorku.ca}
}

\maketitle

\begin{abstract}
Bacteria generally live in complicated structures called biofilms, consisting of communicating bacterial colonies and extracellular polymeric substance (EPS). Since biofilms are related to detrimental effects such as infection or antibiotic resistance in different settings, it is essential to model their formation. In this paper, a stochastic model is proposed for biofilm formation, using bacterial quorum sensing (QS). In this model, the biological processes in the biofilm formation are modeled as a chemical reaction network which includes bacterial reproduction, productions of autoinducer and EPS, and their diffusion. The modified explicit tau-leap simulation algorithm is adapted based on the two-state QS mechanism. Our approach is validated by using the experimental results of \textit{Pseudomonas putida} IsoF bacteria for autoinducer and bacteria concentration. It is also shown that the percentage of EPS in the biofilm increases significantly after the state change in QS, while it decreases before QS is activated. The presented work shows how the biofilm growth can be modeled realistically by using the QS mechanism in stochastic simulations of chemical reactions.
\end{abstract}

\section{Introduction}
Bacteria can attach to surfaces, and while attached, can form sophisticated colonies called biofilms. Briefly, a biofilm is a structure consisting of cooperating bacteria, and substances they produce such as extracellular polymeric substance (EPS). Biofilms can be found in natural, medical and industrial environments \cite{lopez2010biofilms}. For instance, bacteria can form biofilms in underwater parts of vessels, teeth,  and biomedical implants. In vessels, they cause more fuel consumption due to the increased hydrodynamic drag. In human body, they are associated with bacterial infections and antibiotic resistance \cite{rather2021microbial}. Therefore, modeling the growth of biofilms is a significant research subject. 

As for biofilm formation modeling, the proposed methods can be classified as continuum and discrete approaches \cite{mattei2018continuum}. In continuum models, the spatiotemporal spread of the biomass is modeled by deterministic partial differential equations. This biomass is assumed to grow and propagate in the medium based on diffusion and flows. 
In the discrete approach, cellular automaton (CA) models, hybrid differential-discrete CA models and individual based models (IbMs) are employed \cite{mattei2018continuum}. CA models estimate the spread of the biomass generally in a 2-D grid similar to the Game of Life, which was developed by the mathematician John Horton Conway. The biomass is represented as square components in the grid and they spread with a set of simple rules to other squares. In the hybrid models, the biomass growth is modeled with CA, while the nutrient spread is modeled by differential equations. In IbMs, the bacterial cell is considered as the fundamental element. Bacteria can move in any random direction of a set of continuous directions as opposed to the CA models. The diffusion and reaction of substrates are also modeled with differential equations. In addition, bacterial communities employ a cell-to-cell communication mechanism called quorum sensing (QS), which is used to sense their population via emitting autoinducer molecules and make some decisions such as biofilm formation or bioluminescence \cite{mukherjee2019bacterial}. QS is incorporated into the biofilm models as a threshold-based decision mechanism depending on the autoinducer concentration \cite{perez2016mathematical}. While these models are highly complex by considering nutrient consumption and the effect of flows in the medium via coupled partial differential equations, they provide deterministic solutions \cite{ward2001mathematical, chopp2003dependence,janakiraman2009modeling, klapper2010mathematical, vaughan2010influence, frederick2011mathematical, tam2018nutrient, tam2019thin, tam2022thin}.  

Since the biofilm modeling includes communication among bacteria, there are also a few studies about biofilms in molecular communications (MC) literature. In \cite{martins2016using}, a deterministic biofilm disruption model is proposed by constructing a bacterial wall around the biofilm using engineered bacteria. In \cite{martins2018molecular}, a deterministic model in which the communication signals used for QS are jammed by engineered bacteria to prevent biofilm formation is proposed. Furthermore, a stochastic model is proposed in \cite{michelusi2016queuing} for only modeling the QS by using a queuing model which assumes the intracellular autoinducer related processes and bacterial reproduction as stochastic processes. 


\begin{figure*}[t]
\vspace{0.1in}
	\centering
	\scalebox{0.75}{\includegraphics{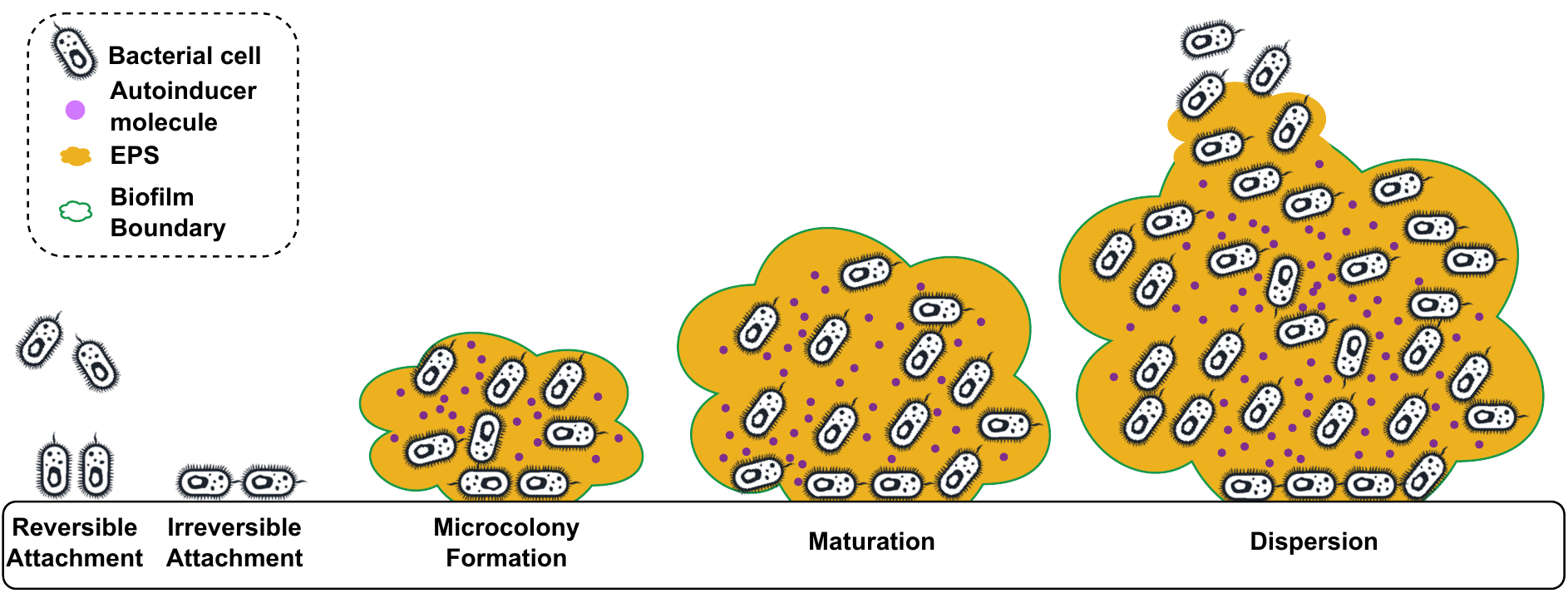}}
	\caption{
		Stages in biofilm formation.
	}
	\label{Stages}
	\vspace{-0.5cm}
\end{figure*}

In the biofilm modeling literature, the CA based approach is dependent on the grid structure and the inclusion of QS is not feasible. In addition, the stochastic effects may be unrealistic due to the uneven cell division and random directions of particles in IbMs \cite{mattei2018continuum}. The deterministic approach based on differential equations are powerful but lack the realistic effects due to the randomness of biological processes in the biofilm. Therefore, an alternative realistic approach which includes randomness and QS is needed for biofilm modeling. 

In this paper, an approach for biofilm formation based on a QS-based decision mechanism and stochastic simulation of biological processes modeled as chemical reactions is proposed. These biological processes during the biofilm growth are modeled as a chemical reaction network (CRN). In this CRN, reproduction of bacteria, EPS production, autoinducer production and its degradation, and the diffusion of the EPS and bacteria are modeled as first order chemical reactions. The QS is modeled as a decision mechanism based on two states via a threshold. In the first state (downregulation), the autoinducer and EPS productions occur with a low rate whereas in the second state (upregulation) these rates are higher. Then, the modified explicit tau-leap stochastic simulation algorithm given in \cite{cao2006efficient} is adapted for the QS-based biofilm formation in a compartmental domain. Numerical results are validated by experimental in vitro results of \textit{Pseudomonas putida} IsoF from the literature for the growth patterns of autoinducer and bacteria concentrations. It is shown that fluctuations in the growth patterns of the biofilm can be more realistically estimated. It is also revealed that while the percentage of the EPS gets lower in the biofilm in the downregulation in time, its percentage in the biofilm increases after the state changes to upregulation due to QS.

The main contribution of this paper is to employ and adapt a stochastic simulation method for a QS and CRN-based biofilm model. While using deterministic solutions of a CRN can give a good estimate for the average behavior of the biofilm, it is not sufficient to realistically model the actual concentrations due to the randomness of the reactions and molecules. Furthermore, deterministic approaches may fail to model the different stable steady states of a CRN \cite{erban2020stochastic}. Thus, our proposed approach enables to model the stochastic effect on the state change of QS, and the corresponding stochastic growth of the biofilm.


\section{Model}

\subsection{Biological Processes in Biofilm Formation}
Biofilms consist of a bacterial colony, polysaccharides, proteins, and nucleic acids such as DNA and RNA \cite{rather2021microbial}. The substance except bacterial cells is defined as EPS which provides a sticky medium for the bacteria to grow and resist to extracellular effects such as antibiotics in the human body. The first step for the biofilm formation is the attachment of some bacteria to a surface as illustrated in Fig. \ref{Stages}. At the initial phase, which is called the reversible attachment, they are loosely adhered to a surface. Then, they make a transition to an irreversible attachment stage where they change their orientation and adhere to the surface permanently. In the next stage, the biofilm starts to be formed by the reproduction of bacteria and the production of EPS which serves as a scaffold for the bacterial colony. The early stage of this growth is called the microcolony formation, while the latter stage is called maturation. After the maturation stage, the biofilm can be partly disrupted due to the external effects in the medium such as fluid flows and they disperse to form other biofilms as shown in Fig. \ref{Stages}. In this paper, we will focus on the growth phases (microcolony formation and maturation) of the biofilm formation which includes bacterial reproduction and EPS production.

\begin{figure}[!b]
	\vspace{-0.5cm}
	\centering
		\scalebox{0.5}{\includegraphics{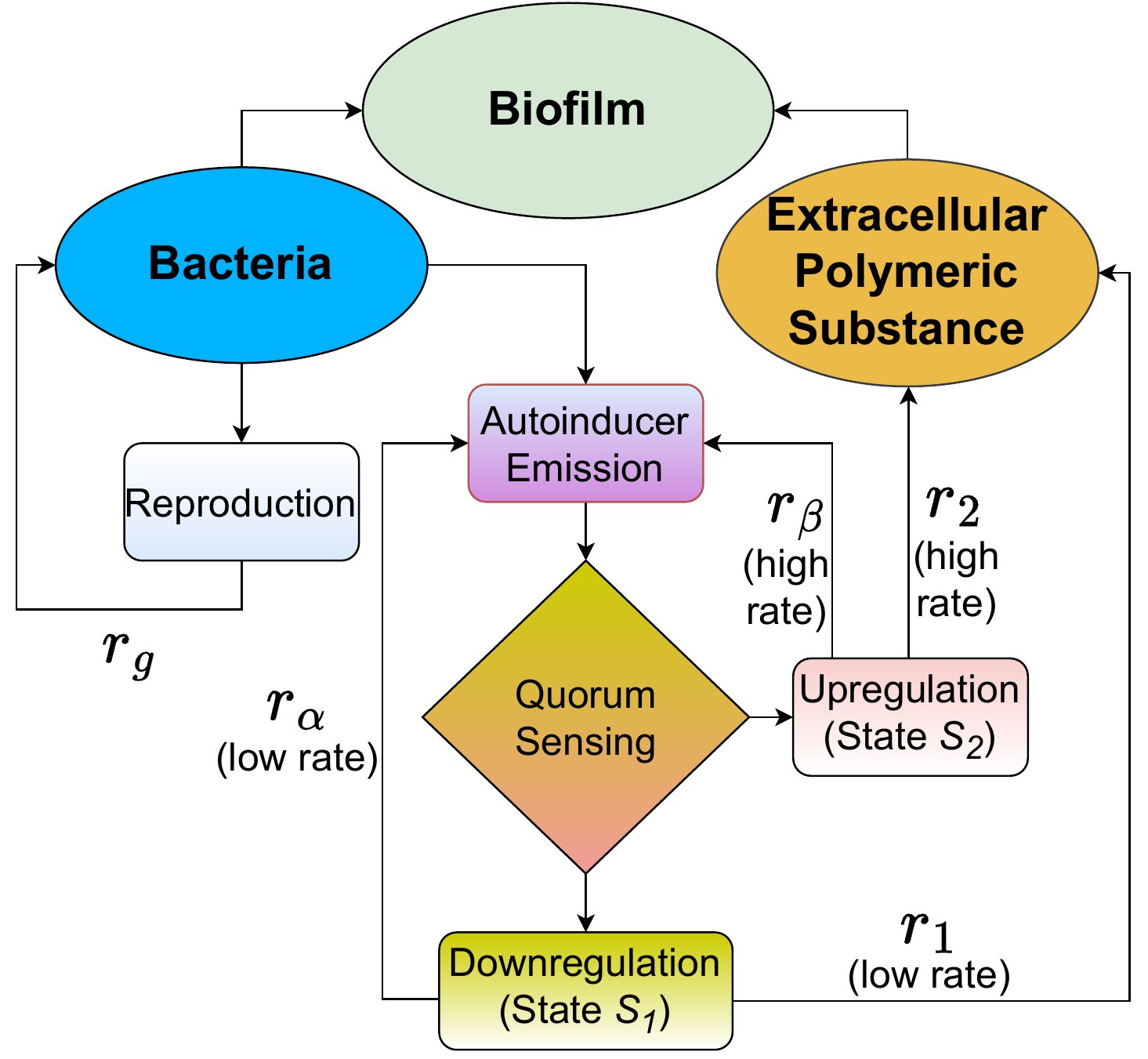}}
	\caption{
		Biological processes in biofilm formation.
	}
	\label{Block}
\end{figure}

As shown in Fig. \ref{Block}, once the bacteria attach to a surface, they start to reproduce, produce EPS, and emit autoinducer molecules which are employed for the communication among bacteria for QS \cite{mukherjee2019bacterial}. During the growth phase, bacteria, EPS and autoinducer molecules diffuse in the medium. In QS, bacteria make a decision according to a threshold of the autoinducer concentration around them to determine downregulation and upregulation states \cite{ward2001mathematical}. In the downregulation state, bacteria produce autoinducers and EPS with lower rates whereas in the upregulation state they produce the autoinducers and EPS with higher rates. In the next section, the system model is elaborated based on these QS-based biological processes.

\subsection{Communication model: Scenario and assumptions} \label{SM}
In order to model the biological processes in biofilm formation, this section elucidates the system model which is based on a CRN in a compartmental domain. First, we begin with our scenario and assumptions.
\vspace{-0.3cm}
\begin{figure}[hbt]
	\centering
	\includegraphics[width=\columnwidth]{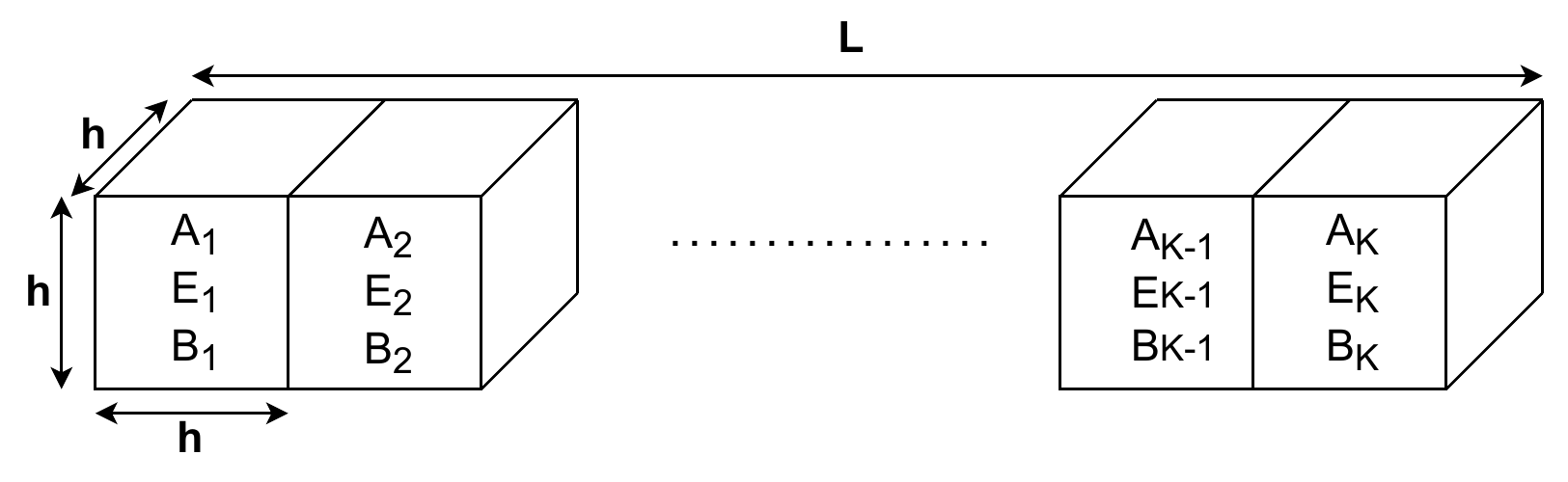}
	\caption{
		Computational domain consisting of K compartments.
	}
	\label{Domain}
	\vspace{-0.2cm}
\end{figure}

The biomass in the biofilm includes the masses of bacteria and EPS for a 1-D medium which consists of $ K $ identical cubic compartments with a length $ h $ on the x-axis. These compartments constitute the computational domain with a total length $L = h\times K$ and total volume $V = h^3 \times K$. As illustrated in Fig. \ref{Domain}, each compartment has a volume of $ h^3 $ and we assume that the concentrations in each compartment is homogeneous. It is also assumed that there are sufficient nutrients in the medium and bacteria do not die. In addition, bacteria and EPS propagate in the medium only via diffusion among compartments with the diffusion coefficient $D$. Since the diffusion coefficient of autoinducer is much greater than $D$, we assume that the autoinducer molecules diffuse almost instantaneously and its concentration becomes homogeneous in the whole domain in accordance with the biological situation \cite{frederick2011mathematical}. Bacteria reproduce and autoinducer molecules degrade with a rate $r_g$ and $r_{\sigma}$, respectively. As given in Fig. \ref{Stages}, if the number of A molecules exceeds a threshold ($\Gamma_{QS}$), then all of the bacteria pass to the upregulation state (state $S_2$). Otherwise, they are in a downregulation state (state $S_1$). In state $S_1$, A is produced with the rate $r_{\alpha}$ and EPS is produced with the rate $r_1$. In state $S_2$, A is produced with the rate $r_{\beta}$ and EPS is produced with the rate $r_2$.

\subsection{Chemical Reaction Network}
We model all of the biological processes summarized in Fig. \ref{Block} as a CRN taking place in each compartment separately. Moreover, diffusion of bacteria and EPS are modeled as chemical reactions between the subsequent compartments \cite{erban2020stochastic}. In this CRN, reactions are coupled so that the number of molecules in one species can affect another reaction. 

The reactions for the $i^{th}$ compartment where  $i = 1, 2,...,K$ are given below by classifying them as reactions according to their states. In these reactions, $A_i$, $B_i$ and $E_i$ represent the autoinducer, bacteria and EPS in the $i^{th}$ compartment, respectively. Moreover, $\emptyset$ shows the chemical species that are of no interest.

\setlength{\belowdisplayskip}{0pt} \setlength{\belowdisplayshortskip}{0pt} 
\setlength{\abovedisplayskip}{0pt} \setlength{\abovedisplayshortskip}{0pt}

\paragraph{State $S_1$}
\begin{equation}
	\ce{\emptyset{} + B_i ->[r_\alpha] A_i + B_i},
	\label{R1a}
\end{equation}

\begin{equation} \label{R3a}
	\ce{\emptyset{} + B_i ->[r_1] B_i + E_i} 		
\end{equation}

\paragraph{State $S_2$} 
\begin{equation}
	\ce{\emptyset{} + B_i ->[r_\beta] A_i + B_i},
	\label{R1b}
\end{equation}

\begin{equation} \label{R3b}
	\ce{\emptyset{} + B_i ->[r_2] B_i + E_i} 		
\end{equation}

\paragraph{All states}
\begin{equation} \label{R2a}
	\ce{A_i ->[r_\sigma] \emptyset{}} 		
\end{equation}

\begin{equation} \label{R4a}
	\ce{\emptyset{} + B_i ->[r_g] B_i + B_i} 		
\end{equation}


\begin{equation} \label{R6}
	\ce{E_1 <=>[d_B][d_B] E_2 <=>[d_B][d_B] \hspace{1cm} ... \hspace{1cm} <=>[d_B][d_B] E_K} 		
\end{equation}

\begin{equation} \label{R7}
	\ce{B_1 <=>[d_B][d_B] B_2 <=>[d_B][d_B] \hspace{1cm} ... \hspace{1cm} <=>[d_B][d_B] B_K} 		
\end{equation}

\vspace{0.2cm}

\setlength{\belowdisplayskip}{5pt} \setlength{\belowdisplayshortskip}{5pt} 
\setlength{\abovedisplayskip}{5pt} \setlength{\abovedisplayshortskip}{5pt}

Reactions (\ref{R1a}) and (\ref{R1b}) give the production of autoinducer molecules depending on the number of bacteria for different states. Similarly, (\ref{R3a}) and (\ref{R3b}) represent the production of EPS related with the number of bacteria for two states. For the reactions (\ref{R1a})-(\ref{R3b}), the state is determined according to the threshold $\Gamma_{QS}$ as given by
\begin{equation} \label{state_check}
	C_A(t) \mathop{\lessgtr}_{S_2}^{S_1} \Gamma_{QS},
\end{equation}
where $C_A(t)$ is the average concentration of the autoinducer molecules in the whole domain. It is  calculated as $C_A(t) = (1/V) \sum_{i=1}^{K} N_{A_i}(t)$ where $N_{A_i}(t)$ is the number of autoinducer molecules in the $i^{th}$ compartment. In addition, (\ref{R2a}) and (\ref{R4a}) correspond to the degradation of autoinducer molecules and reproduction of bacteria, respectively.

While (\ref{R1a})-(\ref{R4a}) take place in the same compartment, (\ref{R6})-(\ref{R7}) show the reactions between compartments to model the diffusion. Furthermore, the reaction rate constants for the diffusion are calculated as $d_B = D/h^2$ \cite{erban2020stochastic}. Next, the simulation method for the numerical solutions of the concentrations in the CRN given in this section is explained.

\section{Stochastic Simulation of the Chemical Reaction Network} \label{SS_CRN}
In this section, we elaborate the simulation method for the CRN given in (\ref{R1a})-(\ref{R7}) to estimate the growth of the biofilm with its components in time and space. In the stochastic approach, the chemical master equation (CME) can be used to determine the probabilities that any species in a CRN have a certain number of molecules \cite{gillespie2007stochastic, egan2022stochastic}. However, it is nontrivial to solve this CME for the CRN given in (\ref{R1a})-(\ref{R7}) and it is even very difficult to write the CME itself. Therefore, stochastic simulation algorithms (SSAs) are employed to solve such realistic systems. In this paper, the modified explicit tau-leap algorithm, which is a fast and accurate SSA to determine the stochastic evolution of the biofilm \cite{cao2006efficient}.


Firstly, some preliminary definitions are made for the modified explicit tau-leap algorithm. Let $\mathbf{X}(t) = \left( A_1(t), E_1(t), B_1(t), ... , A_K(t), E_K(t), B_K(t) \right)$ be the state vector holding the number of molecules for each species of the CRN given in (\ref{R1a})-(\ref{R7}). In addition, $\pmb{\nu}_j = (\nu_{j,1},...,\nu_{j,3K})$ is defined as the state change vector showing the change in the number of species according to the stoichiometric coefficients for the $j^{th}$ reaction. $\mathbf{K}_j(\tau;x,t)$ is defined as the number of times that $j^{th}$ reaction occurs in the time interval $[t,t+\tau)$ when $\mathbf{X}(t) = \mathbf{x}$. $a_j(\mathbf{x}) dt$ gives the probability for the $j^{th}$ reaction to take place in the time interval $[t,t+\tau)$ given that $\mathbf{X}(t) = \mathbf{x}$ and $a_j(\mathbf{x})$ is the propensity function. 

Before starting the steps of the modified explicit tau-leap algorithm, an initial step is added to the simulation to determine the states according to (\ref{state_check}). If the state is determined as $S_1$, then the reactions (\ref{R1a})-(\ref{R3a}) are chosen. If the state is $S_2$, then reactions (\ref{R1b})-(\ref{R3b}) are chosen. Reactions (\ref{R2a})-(\ref{R7}) are considered for both states. Based on these chosen reactions, the propensities are updated. After the initial step for the states,  the modified explicit tau-leap algorithm is given based on the procedure detailed in \cite{cao2006efficient} as follows. 

First, the maximum number of times ($\Lambda_j$) that $j^{th}$ reaction can occur is calculated as
\begin{equation}
	\Lambda_j = \min_{ i \in [1,3K], \nu_{j,i} < 0 }{\left \lfloor \frac{x_i}{|\nu_{j,i}|} \right \rfloor },
\end{equation}
where the brackets show the floor function. Hence, the noncritical reactions, which are not likely to exhaust all molecules in one time step, and critical reactions are determined according to a threshold ($n_c$). If $\Lambda_j < n_c$ and $a_j(\mathbf{x}) > 0$, then this reaction is considered as critical. Then two candidate time steps, i.e., $\tau'$ and $\tau''$ are calculated. $\tau'$ is given by
\begin{equation}
\hspace{-0.4cm}	\tau' = \min_{ i \in I_{rs}}{ \left\{ \frac{\max\{\epsilon,1\}}{|\sum_{j \in J_{ncr}} \nu_{j,i} a_j(\mathbf{x}) |}, \frac{\max\{\epsilon,1\}^2}{|\sum_{j \in J_{ncr}} \nu_{j,i}^2 a_j(\mathbf{x}) |} \right\}  } \hspace{-0.4cm}
\end{equation}
where $I_{rs}$ and $J_{ncr}$ denote the set of indices of all species in CRN and indices of noncritical reactions, respectively, and $\epsilon$ is a bounding parameter. If $\tau' < (10/a_0(\mathbf{x}))$, then the simulation is continued along $100$ time steps with the Gillespie first reaction algorithm and the simulation starts over from the first step. Otherwise, $\tau''$ is calculated as a random variable drawn from an exponential distribution with the mean value $ 1/a_0^c(\mathbf{x}) $ where $a_0^c(\mathbf{x}) $ is the sum of the propensity function values for the critical reactions. 

In the next step, the smaller one of the candidate time steps is chosen as $\tau$. If $\tau'$ is the smaller one, then no critical reaction is executed and $\mathbf{K}_j(\tau;x,t)$ is a Poisson random variable with the rate $a_j(\mathbf{x}) \tau$ for noncritical reactions. Otherwise, only one critical reaction is chosen to occur according to their probabilities calculated via the propensities and $\mathbf{K}_j(\tau;x,t)$ is set in the same way. In the last step, if there is a negative value in $ \mathbf{X}(t) + \sum_{j=1}^{M} \mathbf{K}_j(\tau;x,t) \pmb{\nu}_j $, then the state vector is not updated and $\tau'$ is divided by two and the simulation continues from the step where $\tau' < (10/a_0(\mathbf{x}))$ is checked. Otherwise, the states at each time step are updated as 
\begin{equation} \label{update}
	\mathbf{X}(t+\tau) = \mathbf{X}(t) + \sum_{j=1}^{M} \mathbf{K}_j(\tau;x,t) \pmb{\nu}_j,
\end{equation}
where $M = 8K-4$ is the number of reactions. The results obtained with this algorithm are given in terms of average concentration by dividing the total number of particles for each species in the whole domain to $V$ to compare the results with the experimental in vitro data as given in the next section.

\begin{table}[b]
	\vspace{-0.6cm}
	\centering
	\caption{Simulation parameters}
	\centering\setcellgapes{2pt}\makegapedcells \renewcommand\theadfont{\normalsize\bfseries} 
	\scalebox{0.80}{
		\begin{tabular}{p{70pt}|p{55pt}|p{70pt}|p{55pt}}
			\hline
			\textbf{Parameter}	& \textbf{Value} & \textbf{Parameter}	& \textbf{Value}\\
			\hline \hline
			Specific autoinducer production rate at $S_1$ ($\mu_{\alpha}$) & $ 2.3 \times 10^{-19} $ mol per cell h$ ^{-1} $ \cite{fekete2010dynamic} & Specific autoinducer production rate at $S_2$ ($\mu_{\beta}$) & $ 2.53 \times 10^{-18} $ mol per cell h$ ^{-1} $ \cite{fekete2010dynamic} \\ \hline
			EPS production rate at $S_1$ ($r_1$) & $0.035$ h$ ^{-1} $ \cite{frederick2011mathematical} & EPS production rate at $S_2$ ($r_2$) & $0.35$ h$ ^{-1} $ \cite{frederick2011mathematical} \\ \hline
			Diffusion coefficient ($D$) & $2.78 \times 10^{-10}$ m$^2$ h$ ^{-1} $ \cite{frederick2011mathematical} & Autoinducer degradation rate ($r_{\sigma}$) & $0.0046$ h$ ^{-1} $ \cite{frederick2011mathematical} \\ \hline
			Specific reproduction rate of bacteria ($\mu_g$) & 0.0417 h$ ^{-1} $ \cite{frederick2011mathematical} & QS detection threshold ($\Gamma_{QS}$)&  $3.011 \times 10^{15}$ particles l$^{-1}$ \\ \hline 
			Initial number of autoinducer, bacteria and EPS molecules/population & $ 20 $ & Initial positions of autoinducer, bacteria and EPS molecules/population & $16$th-$17$th compartment \\ \hline
			Number of compartments ($K$) & 32 & Domain length ($L$) & $ 7 $ mm \\ \hline
			Critical reaction threshold ($n_c$) & 10 \cite{cao2006efficient} & Bounding parameter ($\epsilon$) & 0.03 \cite{cao2006efficient}	\\	 \hline
			\hline           
	\end{tabular}
	}
	\label{Sim_parameters}
\end{table}

\section{Numerical Results} \label{NR}
In this section, numerical results for the biofilm formation are given and validated by the experimental in vitro results. The simulation parameters are given in Table \ref{Sim_parameters}. The experimental values in \cite{fekete2010dynamic,frederick2011mathematical} are employed for the production rates of autoinducer, EPS, and bacterial reproduction and autoinducer degradation. For the in vitro results, the experimental results of \textit{Pseudomonas putida} IsoF bacteria which produce N-Acyl homoserine lactones (AHLs) as the autoinducer molecules. Here, $D$ is taken as an average value of the most dilute and dense diffusion coefficient values in \cite{frederick2011mathematical}. The bacterial reproduction rate is calculated as $r_g = \mu_g B_0$ where $B_0$ is the initial number of bacteria and $\mu_g$ is the specific reproduction rate. Similarly, EPS production rates are calculated as $r_\alpha = \mu_{\alpha} N_{av} B_0$ and $r_\beta = \mu_{\beta} N_{av} B_0$ where $N_{av}$ is the Avogadro constant, $\mu_{\alpha}$ and $\mu_{\beta}$ are the specific autoinducer production rates for state $S_1$ and $S_2$, respectively. $\Gamma_{QS}$ is chosen as a small value which is found by multiplying $5$ n mol l$^{-1}$ (in accordance with the biology literature \cite{frederick2011mathematical}) with $N_{av}$ to have a similar QS state change of the experimental in vitro data. The domain length is chosen in the same mm scale as in \cite{frederick2011mathematical}. In addition, critical reaction threshold and bounding parameter values are taken from \cite{cao2006efficient}. 

\begin{figure}[tb]
	\centering
	\includegraphics[width=\columnwidth]{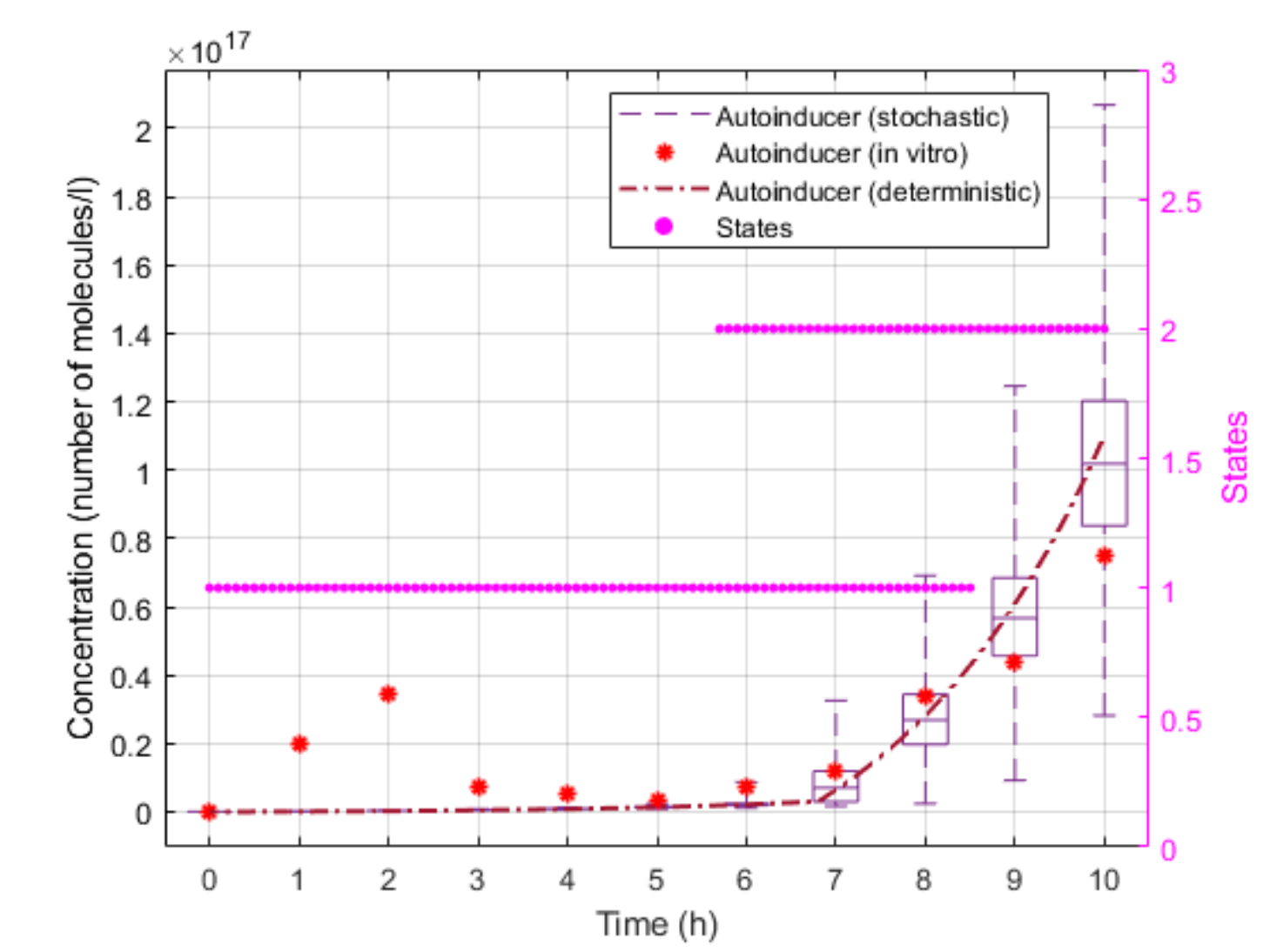} 
	\caption{
		Box plot of average autoinducer concentration and distribution of states.
	}
	\label{Plot_A}
	\vspace{-0.4cm}
\end{figure}
\begin{figure}[!b]
	\vspace{-0.5cm}
	\centering
	\includegraphics[width=\columnwidth]{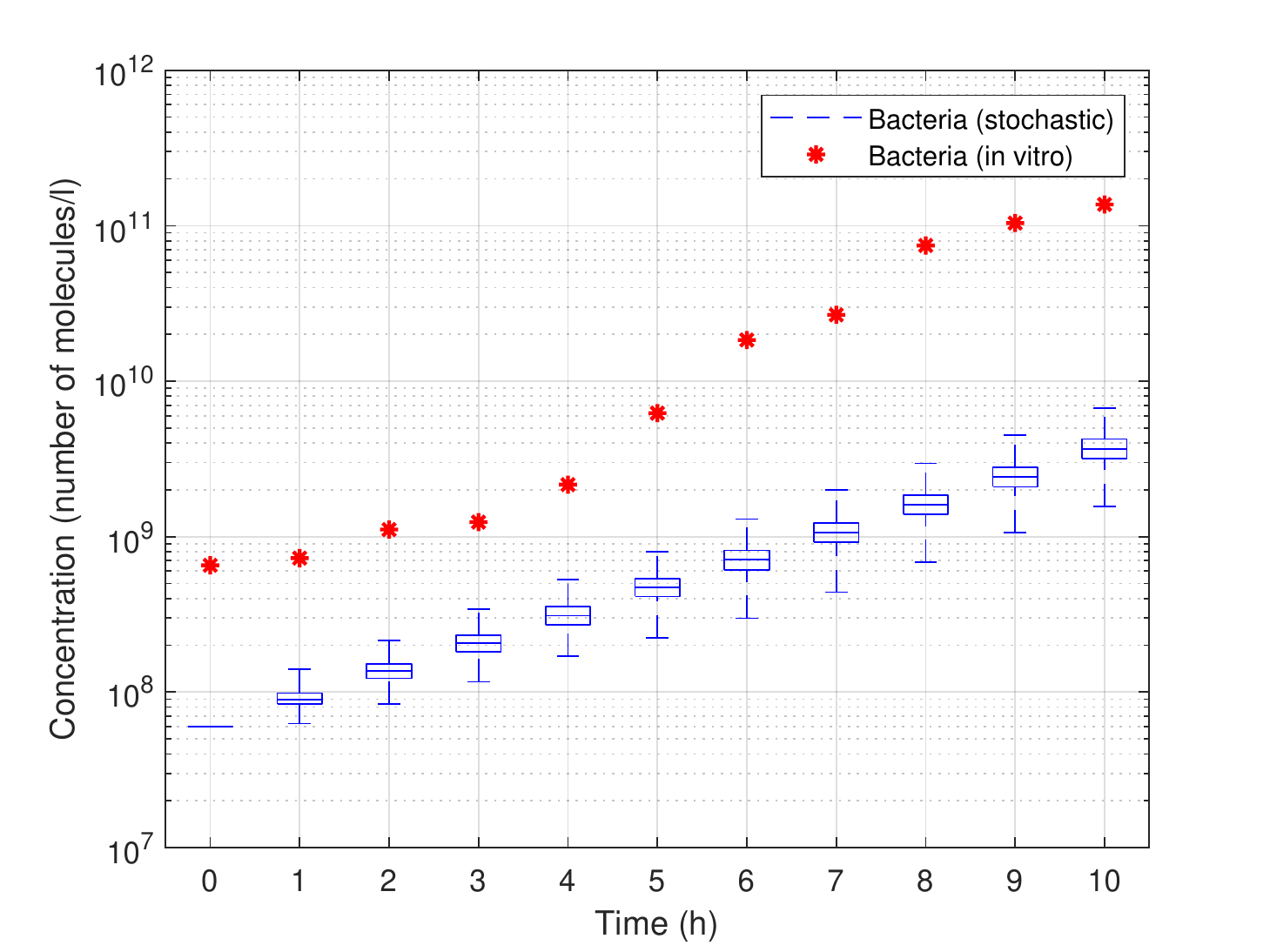} 
	\caption{
		Box plot of average bacteria concentration and distribution of states.
	}
	\label{Plot_B}
\end{figure}

The stochastic simulations are highly time consuming for realistic scenarios such as the biofilm formation lasting for $10$ hours and simulations are performed $1000$ times. The reason for the long simulation time is the large reaction rates, since they cause a small time step as explained in Section \ref{SS_CRN}. In order to be able to finish the simulation in a reasonable period, $r_{\alpha}$, $r_{\beta}$ and $\Gamma_{QS}$ values are downscaled by multiplying with a scaling coefficient ($f_s = 10^{-6}$) initially and the number of autoinducer molecules are upscaled by multiplying with ($1/f_s$) at the end of the simulation. This scaling did not affect the simulation results as verified by the deterministic solution of the CRN given in Fig. \ref{Plot_A}. For the deterministic solution of the CRN, Matlab SimBiology Toolbox is adapted for a variable threshold scenario. For reactions and diffusion, the coupled ordinary differential equations are solved for the corresponding compartments with the first order numerical differentiation formulas \cite{shampine1997matlab}.

\begin{figure}
	\centering
	\includegraphics[width=\columnwidth]{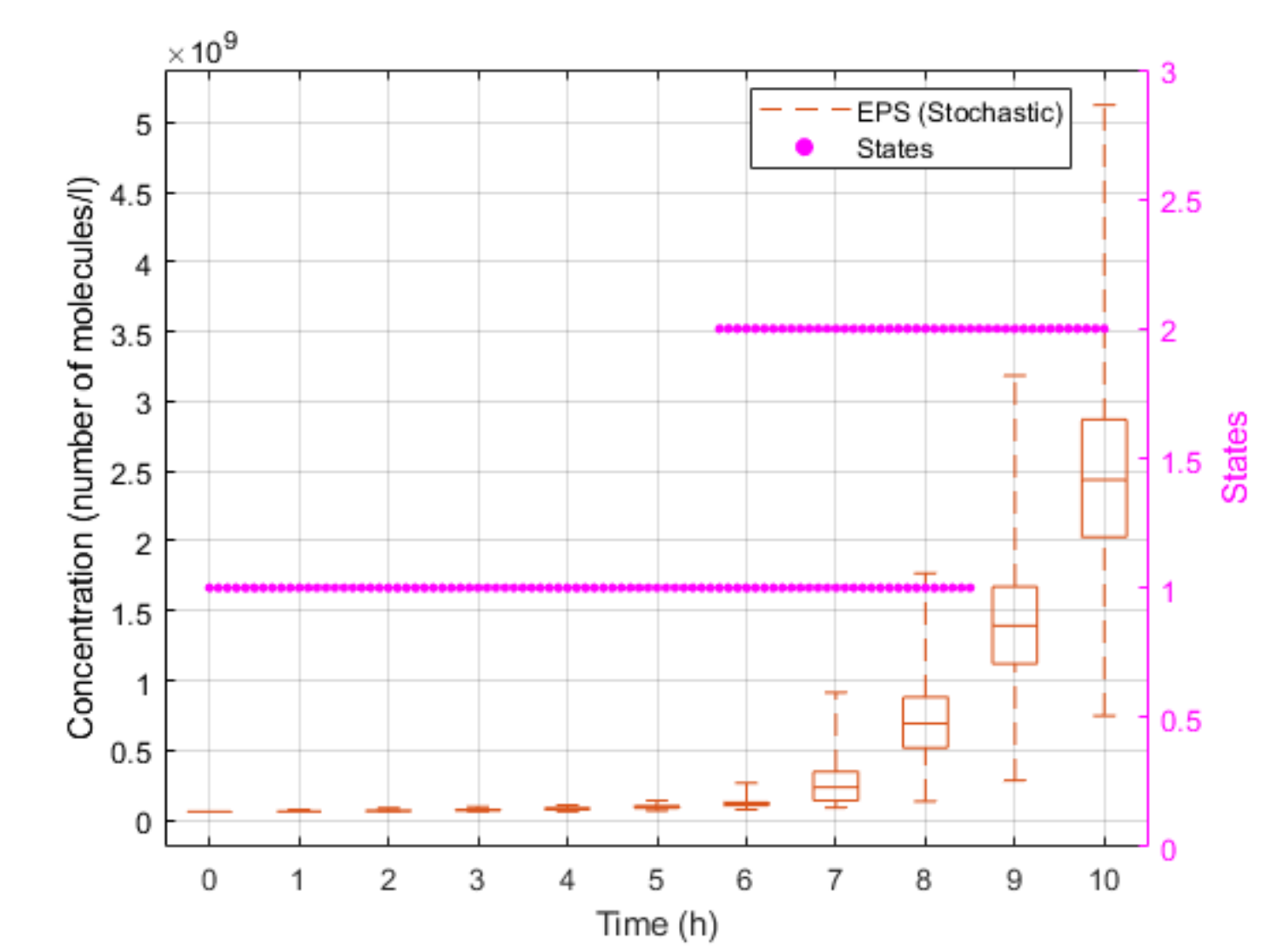} 
	\caption{
		Box plot of average EPS concentration and distribution of states.
	}
	\label{Plot_E}
	\vspace{-0.4cm}
\end{figure}

\begin{figure}[t]
	\centering
	\includegraphics[width=\columnwidth]{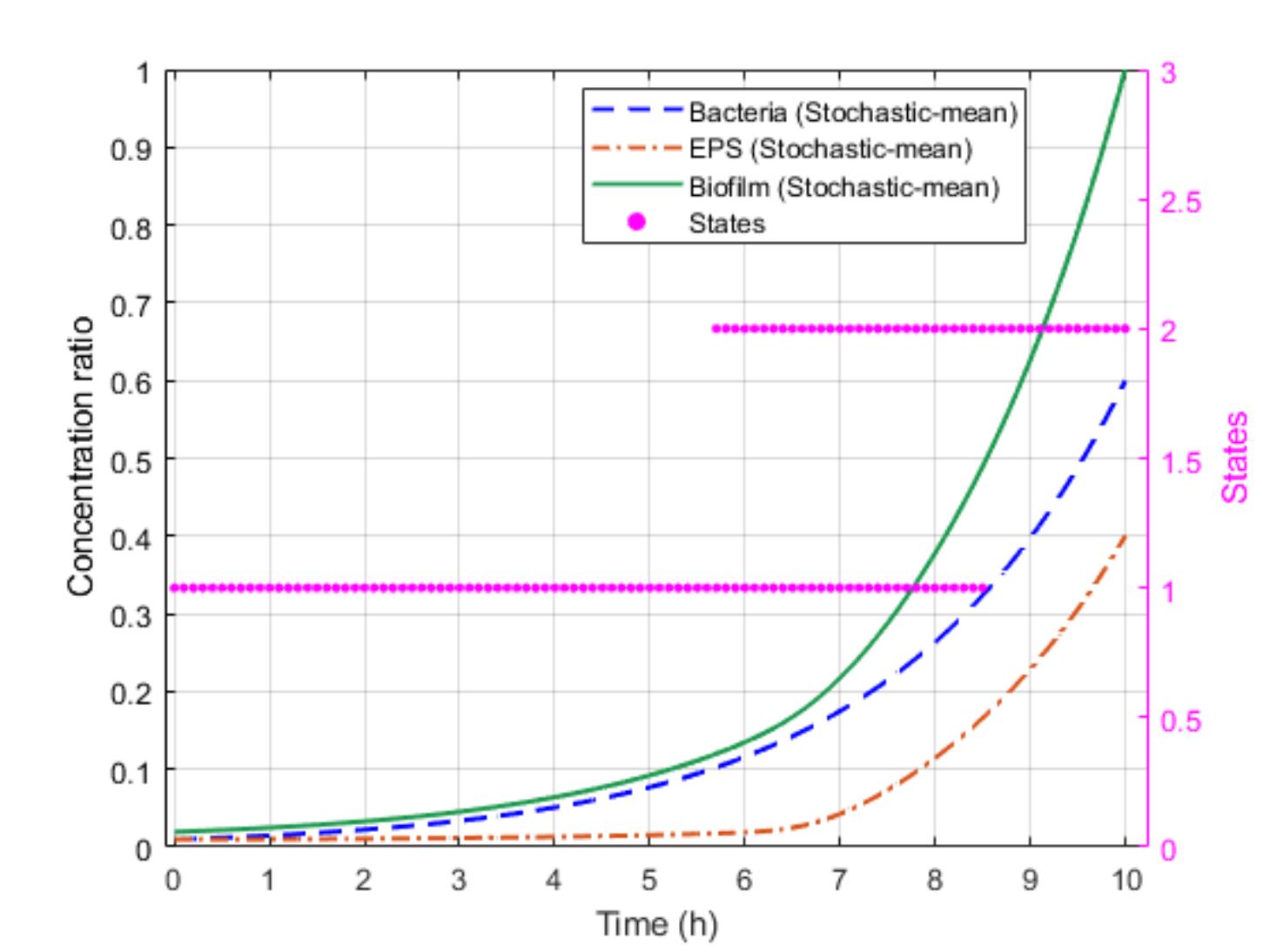} 
	\caption{
		Average concentration ratios of the biofilm and its components with the distribution of states.
	}
	\label{Plot_biofilm_t}
	\vspace{-0.5cm}
\end{figure}

As shown in Fig. \ref{Plot_A}, our proposed model mostly overlaps for the autoinducer concentration, i.e., 3-oxo-C10-HSL which is a type of AHL, with the experimental in vitro results in \cite{fekete2010dynamic}. Here, results of the stochastic simulation is plotted with a box plot, since the distribution of the results are not always Gaussian. The upper and lower bound of the boxes show the $75\%$ and $25\%$ quartiles, respectively, and the whiskers show the minimum and maximum values. Except the irregular fluctuations at the beginning of the in vitro results which are related to the experimental measurement conditions of the baffled flask as stated in \cite{fekete2010dynamic}, the fluctuations are estimated well by the stochastic model. Hence, our model, which actually monitors the effect of communication among bacteria, shows its applicability to estimate the time course of autoinducer concentration. Approximately after the sixth hour, bacteria change their state to a upregulation state ($S_2$) with a higher rated autoinducer production. The distribution of the states are also denoted in Fig. \ref{Plot_A} (as well as Figs. \ref{Plot_E} and \ref{Plot_biofilm_t}) to give an insight and variability of the state changes from $S_1$ to $S_2$ based on the reactions even if the threshold is constant. From this state distribution, it is observed that the detection time for QS, i.e., the time that bacteria change their state to $S_2$, is distributed between $5.5$-$10$ h.

\begin{figure}[b]
	\vspace{-0.5cm}
	\centering
	\includegraphics[width=\columnwidth]{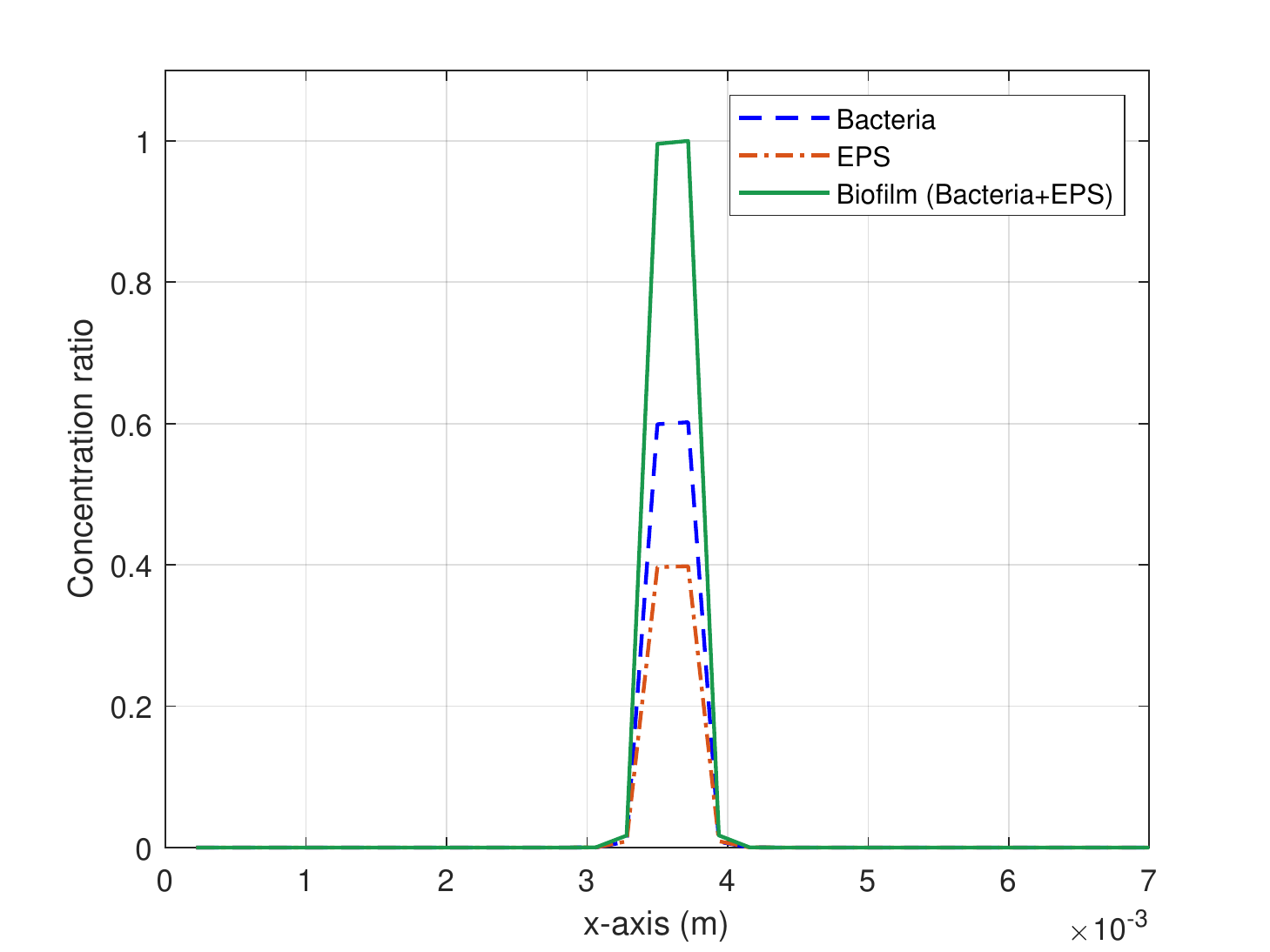} 
	\caption{
		Average concentration ratios of the biofilm and its components on x-axis.
	}
	\label{Plot_biofilm_x}
\end{figure}

The results given as a box plot in Fig. \ref{Plot_B} also shows that our model can estimate the growth pattern of the bacterial population. The relatively small difference is related to the inconsistency of the 1-D simulation and 3-D well-mixed in vitro experimental conditions. Despite this, stochastic modeling shows its advantage by estimating the fluctuations in the concentration of the bacterial population.

Fig. \ref{Plot_E} denotes the box plot of the time course of the EPS concentration in the biofilm. Similar to the autoinducer concentration profile, EPS production boosts after the state change approximately after the sixth hour. In addition, it is observed that the increased EPS production rate due to QS puts the CRN into a more fluctuating state, which approves the necessity of a stochastic solution for the biofilm formation. 

Fig. \ref{Plot_biofilm_x} shows the spatial biofilm formation on the x-axis at the end of the 10 hours simulation. Since the values are normalized according to the maximum concentration,  the relatively small biomass in the regions except the starting point of the biofilm are not clearly visible. This formation shows that biofilm tends to grow firstly in the region the bacteria attaches on the surface.

Furthermore, Fig. \ref{Plot_biofilm_t} is given to compare the roles of the biomass components, i.e., bacteria and EPS, in the growth of the biofilm with their normalized stochastic mean values. Here, it is observed that while the ratio of the bacteria in the biofilm increases from the beginning with respect to EPS, this pattern reverses in favor of EPS after the state change due to the increased rate of EPS production. This observation shows that the communication of bacteria via the EPS production significantly affects the biofilm growth. Surely, this stochastic model can be developed by adding the effect of the nutrients for the biofilm growth and using a 2-D domain for simulation. However, this study aims to show the effect of the communication in the biofilm growth with a stochastic approach, which can lead to research to disrupt the communication and the adverse effects of biofilms.

\section{Conclusion} \label{Conc}
In this paper, a stochastic biofilm model for the biofilm formation based on the bacterial QS is proposed. A CRN abstracting the biological processes in the biofilm formation is employed and simulated with the modified tau-leap stochastic simulation algorithm adapted for QS. It is shown that the proposed stochastic approach agrees with the experimental in vitro results and can estimate the fluctuations in biofilm formation. For the future work, it is planned to investigate the disruption of the communication among bacteria and its effect on the biofilm formation.

\bibliographystyle{ieeetran}

\bibliography{IEEEabrv,ref_fg_biofilm_qs}

\begin{thebibliography}{10}
\providecommand{\url}[1]{#1}
\csname url@samestyle\endcsname
\providecommand{\newblock}{\relax}
\providecommand{\bibinfo}[2]{#2}
\providecommand{\BIBentrySTDinterwordspacing}{\spaceskip=0pt\relax}
\providecommand{\BIBentryALTinterwordstretchfactor}{4}
\providecommand{\BIBentryALTinterwordspacing}{\spaceskip=\fontdimen2\font plus
\BIBentryALTinterwordstretchfactor\fontdimen3\font minus
  \fontdimen4\font\relax}
\providecommand{\BIBforeignlanguage}[2]{{%
\expandafter\ifx\csname l@#1\endcsname\relax
\typeout{** WARNING: IEEEtran.bst: No hyphenation pattern has been}%
\typeout{** loaded for the language `#1'. Using the pattern for}%
\typeout{** the default language instead.}%
\else
\language=\csname l@#1\endcsname
\fi
#2}}
\providecommand{\BIBdecl}{\relax}
\BIBdecl

\bibitem{lopez2010biofilms}
D.~L{\'o}pez, H.~Vlamakis, and R.~Kolter, ``Biofilms,'' \emph{Cold Spring
  Harbor perspectives in biology}, vol.~2, no.~7, p. a000398, 2010.

\bibitem{rather2021microbial}
M.~A. Rather, K.~Gupta, and M.~Mandal, ``Microbial biofilm: formation,
  architecture, antibiotic resistance, and control strategies,''
  \emph{Brazilian Journal of Microbiology}, vol.~52, no.~4, pp. 1701--1718,
  2021.

\bibitem{mattei2018continuum}
M.~Mattei, L.~Frunzo, B.~D’acunto, Y.~Pechaud, F.~Pirozzi, and G.~Esposito,
  ``Continuum and discrete approach in modeling biofilm development and
  structure: a review,'' \emph{Journal of mathematical biology}, vol.~76,
  no.~4, pp. 945--1003, 2018.

\bibitem{mukherjee2019bacterial}
S.~Mukherjee and B.~L. Bassler, ``Bacterial quorum sensing in complex and
  dynamically changing environments,'' \emph{Nature Reviews Microbiology},
  vol.~17, no.~6, pp. 371--382, 2019.

\bibitem{perez2016mathematical}
J.~P{\'e}rez-Vel{\'a}zquez, M.~G{\"o}lgeli, and R.~Garc{\'\i}a-Contreras,
  ``Mathematical modelling of bacterial quorum sensing: a review,''
  \emph{Bulletin of mathematical biology}, vol.~78, no.~8, pp. 1585--1639,
  2016.

\bibitem{ward2001mathematical}
J.~P. Ward, J.~R. King, A.~Koerber, P.~Williams, J.~Croft, and R.~Sockett,
  ``Mathematical modelling of quorum sensing in bacteria,'' \emph{Mathematical
  Medicine and Biology}, vol.~18, no.~3, pp. 263--292, 2001.

\bibitem{chopp2003dependence}
D.~L. Chopp, M.~J. Kirisits, B.~Moran, and M.~R. Parsek, ``The dependence of
  quorum sensing on the depth of a growing biofilm,'' \emph{Bulletin of
  mathematical biology}, vol.~65, no.~6, pp. 1053--1079, 2003.

\bibitem{janakiraman2009modeling}
V.~Janakiraman, D.~Englert, A.~Jayaraman, and H.~Baskaran, ``Modeling growth
  and quorum sensing in biofilms grown in microfluidic chambers,'' \emph{Annals
  of biomedical engineering}, vol.~37, no.~6, pp. 1206--1216, 2009.

\bibitem{klapper2010mathematical}
I.~Klapper and J.~Dockery, ``Mathematical description of microbial biofilms,''
  \emph{SIAM review}, vol.~52, no.~2, pp. 221--265, 2010.

\bibitem{vaughan2010influence}
B.~L. Vaughan, B.~G. Smith, and D.~L. Chopp, ``The influence of fluid flow on
  modeling quorum sensing in bacterial biofilms,'' \emph{Bulletin of
  mathematical biology}, vol.~72, no.~5, pp. 1143--1165, 2010.

\bibitem{frederick2011mathematical}
M.~Frederick, C.~Kuttler, B.~Hense, and H.~Eberl, ``A mathematical model of
  quorum sensing regulated eps production in biofilm communities,''
  \emph{Theor. Biol. and Med. Model.}, vol.~8, no.~1, pp. 1--29, 2011.

\bibitem{tam2018nutrient}
A.~Tam \emph{et~al.}, ``Nutrient-limited growth with non-linear cell diffusion
  as a mechanism for floral pattern formation in yeast biofilms,''
  \emph{Journal of Theoretical Biology}, vol. 448, pp. 122--141, 2018.

\bibitem{tam2019thin}
A.~Tam, J.~E.~F. Green, S.~Balasuriya, E.~L. Tek, J.~M. Gardner, J.~F.
  Sundstrom, V.~Jiranek, and B.~J. Binder, ``A thin-film extensional flow model
  for biofilm expansion by sliding motility,'' \emph{Proceedings of the Royal
  Society A}, vol. 475, no. 2229, p. 20190175, 2019.

\bibitem{tam2022thin}
A.~K. Tam, B.~Harding, J.~E.~F. Green, S.~Balasuriya, and B.~J. Binder,
  ``Thin-film lubrication model for biofilm expansion under strong adhesion,''
  \emph{Physical Review E}, vol. 105, no.~1, p. 014408, 2022.

\bibitem{martins2016using}
D.~P. Martins, M.~T. Barros, and S.~Balasubramaniam, ``Using competing
  bacterial communication to disassemble biofilms,'' in \emph{Proc. of the 3rd
  ACM Int. Conf. on Nanoscale Comput. and Commun.}, 2016, pp. 1--6.

\bibitem{martins2018molecular}
D.~P. Martins, K.~Leetanasaksakul, M.~T. Barros, A.~Thamchaipenet, W.~Donnelly,
  and S.~Balasubramaniam, ``Molecular communications pulse-based jamming model
  for bacterial biofilm suppression,'' \emph{IEEE transactions on
  nanobioscience}, vol.~17, no.~4, pp. 533--542, 2018.

\bibitem{michelusi2016queuing}
N.~Michelusi, J.~Boedicker, M.~Y. El-Naggar, and U.~Mitra, ``Queuing models for
  abstracting interactions in bacterial communities,'' \emph{IEEE J. on Sel.
  Areas in Commun.}, vol.~34, no.~3, pp. 584--599, 2016.

\bibitem{cao2006efficient}
Y.~Cao, D.~T. Gillespie, and L.~R. Petzold, ``Efficient step size selection for
  the tau-leaping simulation method,'' \emph{The Journal of chemical physics},
  vol. 124, no.~4, p. 044109, 2006.

\bibitem{erban2020stochastic}
R.~Erban and S.~J. Chapman, \emph{Stochastic Modelling of Reaction-Diffusion
  Processes}.\hskip 1em plus 0.5em minus 0.4em\relax Cambridge University
  Press, 2020, vol.~60.

\bibitem{gillespie2007stochastic}
D.~T. Gillespie, ``Stochastic simulation of chemical kinetics,'' \emph{Annual
  review of physical chemistry}, vol.~58, no.~1, pp. 35--55, 2007.

\bibitem{egan2022stochastic}
M.~Egan, B.~C. Akdeniz, and B.~Q. Tang, ``Stochastic reaction and diffusion
  systems in molecular communications: Recent results and open problems,''
  \emph{Digital Signal Processing}, vol. 124, p. 103117, 2022.

\bibitem{fekete2010dynamic}
A.~Fekete \emph{et~al.}, ``Dynamic regulation of n-acyl-homoserine lactone
  production and degradation in pseudomonas putida isof,'' \emph{FEMS
  Microbiology Ecology}, vol.~72, no.~1, pp. 22--34, 2010.

\bibitem{shampine1997matlab}
L.~F. Shampine and M.~W. Reichelt, ``The matlab ode suite,'' \emph{SIAM journal
  on scientific computing}, vol.~18, no.~1, pp. 1--22, 1997.

\end{thebibliography}

\end{document}